\documentstyle[12pt,epsf]{article}
\newcommand{\la}[1]{\label{#1}}
\newcommand{\lsim}{\,{\buildrel < \over {_\sim}}\,}
\newcommand{\gsim}{\,{\buildrel > \over {_\sim}}\,}

\begin{document}

\begin{titlepage}
\begin{flushright}
CERN-TH/97-97\\
nucl-th/9705027\\
\end{flushright}
\begin{centering}
\vfill

{\bf MINIJETS IN ULTRARELATIVISTIC HEAVY ION COLLISIONS}
 
{\bf AT FUTURE COLLIDERS}
 
 \vspace{0.5cm}
 K.J. Eskola\footnote{kari.eskola@cern.ch},
\vspace{1cm}

{\em  CERN/TH, CH-1211 Geneva 23, Switzerland\\}

{\em  Department of Physics,
P.O. Box 9, 00014 University of Helsinki, Finland\\}
 \vspace{0.3cm}
 
\vspace{1cm}
{\bf Abstract}

\end{centering}

The role of minijet production as initial conditions for QGP production 
at $\tau\sim0.1$ fm$/c$ in nuclear collisions  at the LHC and RHIC 
energies is discussed. 
 
\vspace{0.3cm}\noindent

\vfill
\noindent
CERN-TH/97-97\\
May 1997
 
\end{titlepage}

The main goal of ultrarelativistic heavy ion collisions (URHIC) is to 
create strongly interacting condensed elementary particle matter, 
quark--gluon plasma (QGP), and to study thermodynamics of this new phase of
matter. The existence of the QGP phase has been proved by {\it ab initio} 
calculations in lattice QCD \cite{Laermann}. 

High energy particle physics usually aims at producing as few 
particles  as possible in order to create a simple system 
for the measurements.  The URHIC also aim at a simple system
but by producing as {\it many} QCD quanta as possible.
In hadronic collisions the produced final state is  dilute, 
further interactions negligible, and the system essentially freely 
streaming. In URHIC, one would ideally produce a system with a 
{\it maximal} amount of final state interactions, {\it i.e.} a 
large enough thermal parton system with negligible mean free paths.

From the point of view of QGP formation, as high cms-energies 
$\sqrt s$ and as large nuclei  $A$ as possible are preferable. 
So far, the most energetic nucleus--nucleus collisions 
have been the Au--Au collisions at $\sqrt s = 5$ $A$GeV at the 
AGS\footnote{AGS = Alternating Gradient Synchrotron} in the 
Brookhaven National Laboratory (BNL), and Pb--Pb collisions at 
$\sqrt s = 17.6$ $A$GeV at the SPS\footnote{SPS = Super Proton Synchrotron} 
in CERN. A very strong $J/\Psi$-suppression observed in the central 
Pb--Pb collisions at the SPS by the NA50 collaboration \cite{NA50} 
suggests that the QGP phase may have already been reached at the
SPS, and,  more importantly, observed.  

In future heavy ion collisions, Au--Au at $\sqrt s=200$ $A$GeV 
at the BNL RHIC\footnote{RHIC = Relativistic Heavy  Ion Collider}, 
and Pb--Pb at $\sqrt s=5500$ $A$GeV at the CERN LHC/ALICE\footnote
{LHC = Large Hadron Collider; 
ALICE = A Large Ion Collider Experiment}, 
initial energy densities far beyond the critical densities  of  
QGP formation will be produced. It is also expected that the  
system in the  central rapidity region of the RHIC 
and LHC nuclear collisions will stay in the QGP phase for several  
fm/$c$, thus giving better  possibilities  for observing 
characteristic  signals of the QGP like $J/\Psi$-suppression, 
thermal production of dileptons and photons,  strangeness enhancement, 
collective flow, disoriented chiral condensates, {\it etc.}
For a review, see Ref. \cite{Harris}.

Particle and transverse energy production in the central rapidity 
region of URHIC can be treated as a combination of perturbative 
(hard and semihard) parton production and  non-perturbative (soft) 
particle production.  By ``hard processes'' one usually means 
clearly perturbative processes with momentum or mass scales of 
the order of several or tens of GeV,  while ``semihard'' here 
refers to QCD-processes where partons with transverse momenta of 
a few GeV, ``minijets,'' are produced.  In this article I will 
discuss the role of perturbative parton production as the early 
initial conditions for the QGP \cite{KLL,BM87,EKL} in URHIC 
at LHC and RHIC. In particular, some recent results of initial parton 
production \cite{EK96,EKR} are reviewed in Secs. 2 and 3. I will also 
briefly discuss other approaches \cite{ELR96,McLV} in Sec. 4, and  
further evolution of the QGP in Sec. 5.

By definition,  the semihard processes lie at the border-line of 
hard and soft  physics.  Real hadronic jets have been observed in  
$p\bar p$ collisions from $p_{\rm T} \gsim 5$ GeV \cite{UA188} up 
to $p_{\rm T} \sim 440$ GeV \cite{TEVA}. The minijets with 
$p_{\rm T}\sim 1...2$ GeV are part of the underlying  event rather 
than distinguishable jets\footnote{In this sense  ``minijet'' is not 
a very good name for a semihard parton; it is not a ``jet'' at all.}. 
In nuclear collisions, where thousands  (hundreds) of minijets are 
expected to be produced  in the central rapidity unit at the LHC (RHIC) 
\cite{KLL,EKL,EKR}, detection of individual semihard partons becomes 
impossible.

Even though the semihard partons are not observed as jets, 
their production should fall well within the scope of 
perturbative  QCD (pQCD) since 
$p_{\rm T}\gg\Lambda_{\rm QCD}\sim 200$ MeV.
Ultimately,  to how low values of $p_{\rm T}$ one can push 
the validity of pQCD, is a question of convergence of the QCD 
perturbation series.  The recent results from 
HERA\footnote{HERA = Hadron Electron Ring Anlage} at DESY 
\cite{HERAlatest, HERArecent} indicate that the  behaviour of 
parton densities can be explained within  pQCD down to the low 
scales $Q^2\sim 1$ GeV$^2$. This supports  applicability of pQCD 
to semihard parton production as well. 

In the division into semihard and soft particle production,
the key feature is that one is able to {\it compute} the semihard parton 
production from pQCD, given that the parton distributions of colliding 
hadrons and nuclei are known.  On top of this, the non-perturbative 
particle  production can be modelled {\it e.g.}  through strings  
\cite{STRINGS,VENUS,FRITIOF,HIJING}, or in URHIC perhaps also through 
a decaying strong  background  colour field \cite{KM85,EG93}. 
With increasing energies however,  the semihard QCD-processes are 
expected to become increasingly  important, particularly in URHIC. 
This is due to the following  reasons:
\bigskip

\begin{itemize} 

\item
With increasing cms-energies, the events tend to become more ``jetty''
or, rather, more ``minijetty'', and in $p\bar p$ and $pp$ collisions 
the  rapid rise of the total and inelastic cross sections  with the 
$\sqrt s$ can be explained by the copious production of semihard 
partons with transverse  momenta $p_{\rm T}\ge p_0\sim 1...2$ GeV 
\cite{EIKONAL,XNW91}. Perturbative parton production in $AA$ collisions 
scales as  $\sim A^{4/3}$ \cite{KLL,EKL},  while the soft component 
scales more  like $\sim A$, so for large  nuclei the importance of 
semihard partons should be further increased. 

\item
In the deep inelastic $ep$ scatterings (DIS) at HERA it is observed 
that the  structure function $F_2^p(x,Q^2)$ has a steep rise at small 
values of  Bjorken $x$, at $x\lsim 0.01$, persisting down to scales 
$Q\sim 1$ GeV \cite{HERA93,HERArecent,HERAlatest}.  The 
quark--antiquark sea  is generated by emission from the gluons, 
so the gluon distributions  have this rapid rise as well. 
In minijet production in the central rapidities, the dominant gluonic 
processes at  $p_{\rm T}\sim 2$~GeV probe the parton distributions 
typically at fractional momenta  
$x\sim 2p_{\rm T}/\sqrt s\sim 7\times 10^{-4}$ in the LHC nuclear 
collisions. For the RHIC, where typically $x\sim 0.01...0.02$, the rise of 
the parton distributions will not cause such a big effect at the 
central rapidities.

\item 
Time scale for producing partons and transverse energy into 
the  central rapidity region by semihard collisions is short, typically  
$\tau_{\rm h}\sim 1/p_0\sim 0.1$ fm$/c$, where $p_0\sim 2$ GeV 
is the smallest  transverse momentum included in the computation. 
The soft processes are completed  at later stages of the collision, at 
$\tau_{\rm s}\sim 1/\Lambda_{\rm QCD}\sim 1$ fm$/c$. 
If the density  of partons produced during the hard and semihard stages 
of the heavy ion collision becomes high enough - there are indications 
that it will - fusions start to occur, and a saturation in the initial 
parton production can take place in the 
perturbative region  \cite{BM87,GLRpr,EMW,EK96}. 
As a result, softer particle production will be 
screened.  The fortunate consequence of this is that a larger part of 
transverse energy production in the central rapidities is computable 
from pQCD  at higher energies and the relative contribution from soft 
collisions with $p_{\rm T}\lsim 2$ GeV becomes smaller. Typically, 
the expectation is that at the SPS the soft component clearly dominates, 
and at the LHC the semihard component is the dominant one 
\cite{EKL,XNW91,HIJING}. At the RHIC both components should be  
taken into account. 

\end{itemize}

The importance of copious semihard parton production in an early 
formation  of the QGP in URHIC was first addressed  almost 10 years 
ago, by Blaizot  and  Mueller \cite{BM87},  and by Kajantie, Lindfors 
and Landshoff  \cite{KLL}. In terms of eikonal models for hadronic 
interactions, minijet production had already been emphasized some years 
before, in Refs. \cite{EIKONAL}, to explain the rise in the total 
and inelastic cross sections of $p\bar p$ cross sections beyond the 
CERN ISR energy range.  Models including both strings and semihard 
parton production were introduced  around the same time \cite{SZ87}. 
The work \cite{KLL} was later improved by including partonic cross 
sections  in lowest order (LO) pQCD \cite{EKL} and with a better 
treatment of the rapidity acceptance. Consequences of the HERA results 
\cite{HERA93}  for minijet production in URHIC were studied in 
\cite{EKR}. Minijet production in $AA$ collisions has also been 
studied by Calucci and Treleani from the beginning of the 90's 
\cite{CALUCCI}, and later by Xiong and Shuryak \cite{XIONG}.
In the end of the 80's, systematic efforts began to construct 
event-generators for URHIC, which would be based on the 
increasingly important pQCD component.  As a result, HIJING by 
X.-N. Wang and Gyulassy \cite{HIJING}, and, Parton Cascade Model by 
Geiger and M\"uller \cite{PCM} were constructed. Also other event 
generators have options for simulating the URHIC at the RHIC and LHC 
energies \cite{FRITIOF,DTUNUC,VENUS}.  

Treatment of semihard QCD-scatterings is usually based on collinear 
factorization, where the  hard parton--parton scatterings are factorized 
from the universal parton  distributions at a perturbative scale
$\sim p_{\rm T}\gg\Lambda_{\rm QCD}$. The parton distributions then 
contain the non-perturbative experimental input. A novel approach to 
semihard parton production,  not based on the collinear factorization 
nor independent scatterings, but on a consideration of a classical 
gluon field, has been suggested by McLerran and Venugopalan in 1994 
\cite {McLV}. Not based on collinear factorization either, 
and perhaps directly related to the gluon field approach, 
minijet production in a BFKL approach was studied recently \cite{ELR96}.
These topics will be discussed in Sec. 4.

\section{Minijet production in AA collisions}

The idea of multiple production of semihard gluons and quarks in $pp$ 
and especially in $AA$ collisions is based on a picture of independent 
binary parton--parton collisions. The key quantity is the integrated 
jet cross section:
\begin{equation}
\sigma_{\rm jet}(\sqrt s, p_0,\Delta y) = 
\frac{1}{2}\int_{{p_0^2, \Delta y}} 
        dp_{\rm T}^2dy_1dy_2\sum_{{ijkl=}\atop{q,\bar q,g}}
        \int dy_2\, x_1f_{i/N}(x_1,Q) \, x_2f_{j/N}(x_2,Q)
        {d\hat\sigma\over d\hat t}^{ij\rightarrow kl}
        {\hskip-7mm}(\hat s,\hat t,\hat u),
\label{sigmajet}
\end{equation}
where $x_{1,2}$ are fractional momenta of the incoming partons 
$i$ and $j$. Parton-level quantities are indicated by hats,
and $f_{i/N}(x,Q)$ are the parton distributions in $N$ 
($=p,A$) at a factorization scale, chosen as $Q=p_{\rm T}$.
The normalization factor 2 comes from the fact that, in the lowest 
order (LO) pQCD, there are two partons produced in each semihard 
subcollision. In the eikonal models for $pp$ collisions 
\cite{EIKONAL,XNW91} the ratio $\sigma_{\rm jet}/\sigma_{\rm inelastic}$ 
can be interpreted as the average number of semihard collisions in one 
inelastic event. The explicit numbers I will be quoting in the following 
\cite{EK96} are obtained with the GRV-LO parton distributions \cite{GRV94} 
exhibiting a small-$x$ rise similar to that in the HERA data. More detailed 
formulation can be found in Refs. \cite{KLL,EK96}, and a numerical evaluation 
of Eq. (\ref{sigmajet}) with other parton distributions in Ref. \cite{EKR}.

The above formula is defined in the lowest order, 
$d\hat\sigma/d\hat t\sim\alpha_{\rm s}^2$. Often a constant factor 
$K\sim 2$ is used to simulate the effects of next-to-leading-order (NLO) 
terms.  Studies of the NLO jet cross section $d\sigma/(dp_{\rm T}dy)$ 
\cite{EKS} show that (with a scale choice $Q=p_{\rm T}$ and with a jet 
size $R\sim 1$) this may be a reasonable approximation \cite{EWatHPC}. 
Strictly speaking, however, a theoretical $K$-factor can only be defined 
for quantities where a well-defined, {\it infrared-safe} measurement 
function can be applied \cite{EKS}. For $E_{\rm T}$-production in nuclear 
collisions, an acceptance window in the whole central rapidity unit 
defines such a function, but for this acceptance criterion, and for 
$p_{\rm T}\sim 2$ GeV, the exact NLO contribution has not been considered 
yet. For consistency reasons, however, there is  no $K$-factor included 
in the explicit results  I will discuss here. 

A first estimate of the average number of produced semihard partons in 
a rapidity window $\Delta y$ with $p_{\rm T}\ge p_0$ in an $AA$ collision 
at a fixed impact parameter ${\bf b}$ can be obtained as \cite{KLL}
\begin{equation}
\bar N_{AA}({\bf b},\sqrt s,p_0,\Delta y) = 
2T_{AA}({\bf b})\sigma_{\rm jet}(\sqrt s,p_0,\Delta y),
\label{NAA}
\end{equation}
and the average transverse energy carried by these partons as \cite{KLL}
\begin{equation}
\bar E_{\rm T}^{AA} ({\bf b},\sqrt s,p_0,\Delta y) = T_{AA}({\bf b})
\sigma_{\rm jet}(\sqrt s,p_0)\langle E_{\rm T}\rangle_{\Delta y},
\label{ET}
\end{equation}
where $T_{AA}({\bf b})$ is the nuclear overlap function.
The normalization is $\int d^2{\bf b} T_{AA}({\bf b}) = A^2$,
and since $T_{AA}\sim  A^{4/3}$, it describes the typical scaling of hard 
processes in nuclear collisions. For large nuclei  with Woods--Saxon 
nuclear densities,  $T_{AA}({\bf 0})\approx A^2/(\pi R_A^2)$. 
The acceptance criterion $\Delta y$ will be $|y|\le 0.5$, 
and corresponding cuts will be made for $y_1$ and $y_2$ \cite{EKL}. 
In Eqs. (\ref{NAA}) and (\ref{ET}) above, $T_{AA}({\bf b})\sigma_{\rm jet}$ 
is the average number of semihard collisions and 
$\langle E_{\rm T}\rangle_{\Delta y}$ is the  average transverse 
energy carried  by the partons produced in each of these collisions
into $\Delta y$. Parton-flavour decomposition and  rapidity 
distributions can be found in \cite{EK96}.

In Figs. \ref{results}, the integrated jet cross sections and the first 
$E_{\rm T}$-moments for $\sqrt s=5500$ and 200 GeV are shown as functions 
of the smallest transverse momentum $p_0$ included in the computation. 
These quantities naturally depend strongly on the  choice of $p_0$ 
since the subcross sections diverge as 
$d\hat\sigma/d\hat t\sim p_{\rm T}^{-4}$
when $p_{\rm T}\rightarrow0$. Since $p_0$ is a parameter that decides the 
division between soft and hard physics, by definition some phenomenology 
is needed to fix its value. One way to have control over $p_0$ is to 
study an eikonal model \cite{XNW91}, where the lower limit of $p_0$ 
(and simultaneously a possible $K$-factor) is controlled by the rise 
of the cross sections. In addition, by fitting  measured charged 
particle  spectra in hadronic collisions at high energies, it is possible
to extract a value for $p_0$  in connection with a string model \cite{SZ87}. 
Also, convoluting the partonic cross sections with fragmentation 
functions of each parton flavour  into charged pions and kaons 
\cite{BORZUMATI} gives  a simultaneous handle (in the LO) on the 
$K$-factor at large $p_{\rm T}$ and on $p_0$ at the small $p_{\rm T}$ 
part of  the charged particle $p_{\rm T}$-spectra. These procedures 
result in values $p_0\sim 2$ GeV.

\begin{figure}[tb]
\vspace{4cm}
\centerline{\hspace*{0cm} \epsfxsize=11cm\epsfbox{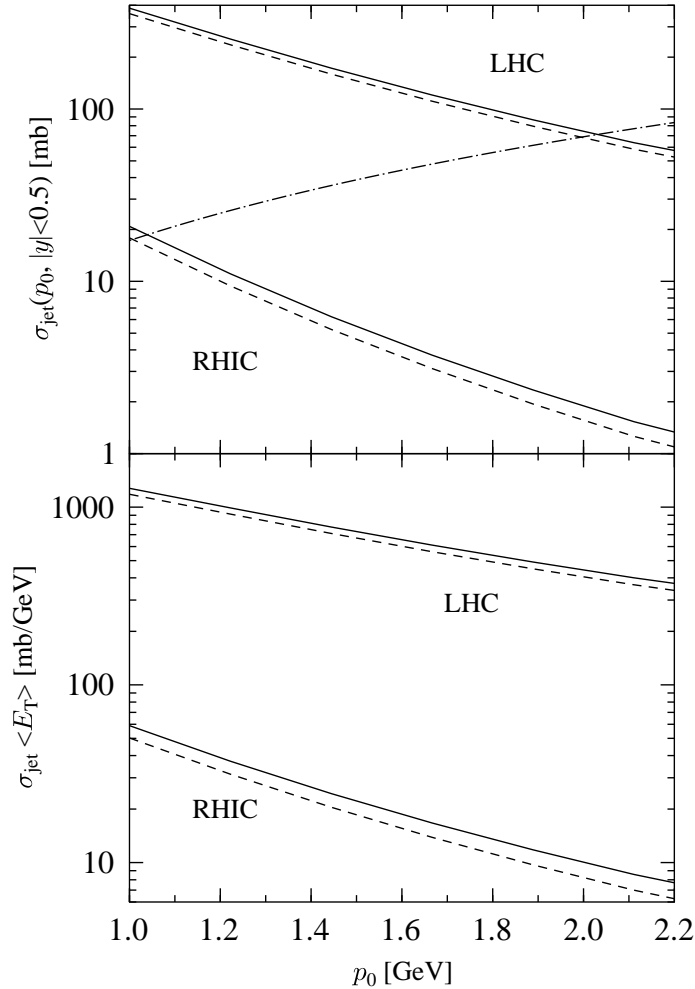}}
\vspace{-2cm}
\caption[a]{ 
{\small The integrated minijet cross section 
$\sigma_{\rm jet}(p_0,\sqrt s,|y|\le0.5)$ 
and the first $E_{\rm T}$-moment 
$\sigma_{\rm jet}(\sqrt s,p_0)\langle E_{\rm T}\rangle_{\Delta y}$ 
as functions of $p_0$ at $|y|\le0.5$ 
at LHC  ($\sqrt s=5500$ GeV) and RHIC ($\sqrt s=200$ GeV)
from Eqs. (\ref{NAA}) and (\ref{ET}).
The dashed curves are the gluon contributions, the dot-dashed curve
shows the transverse saturation limit for $A=208$. Shadowing is not 
included and $K=1$.}
}
\la{results}
\end{figure}

In URHIC then, if sufficiently many partons (mainly gluons) are produced in 
the very beginning of the collision, the partons (within $\Delta y$) 
start to overpopulate the available nuclear transverse area $\sim \pi R_A^2$,
and final-state fusions become important \cite{BM87,GLRpr}. In this case,
further production of partons, especially the transverse energy production,
becomes screened. Let us assign an effective transverse area (uncertainty) 
$\pi/p_{\rm T}^2$ for each parton produced within $|y|\le 0.5$. Since the 
partons dominantly have $p_{\rm T}\sim p_0$, we can estimate \cite{EK96} 
that a  saturation in  the semihard parton production in URHIC should 
happen when $ N_{AA}(p_0,\sqrt s,|y|<0.5)\pi/p_0^2 \gsim \pi R_A^2$, 
{\it i.e.} when
\begin{equation}
\sigma_{\rm jet}(\sqrt s,p_0,|y|\le0.5) \gsim
\frac{R_A^2 p_0^2}{2 T_{AA}({\bf 0})} \sim A^{-2/3} p_0^2,
\label{saturation}
\end{equation}
so the larger the nucleus is, the earlier the saturation will occur.
The saturation value is plotted in Fig.\ref{results} for $A=208$. 
The saturation of the minijet production cross section can be expected 
near $p_{\rm T}\sim 2$ GeV for the LHC and near $p_{\rm T}\sim 1$ GeV 
for the RHIC.  Therefore, for the LHC, the bulk of transverse energy in the 
central rapidity unit is expected to come from the semihard processes alone, 
so $p_0=2$ GeV is a reasonable choice for the LHC. The eikonal model and 
the fragmentation function analysis indicate that a choice $p_0=1$ GeV 
would result in an overestimate of the (inelastic) cross  sections and 
the charged particle $p_{\rm T}$-spectra. Instead of trying to fit this 
value any better for the RHIC, we will consider here the initial conditions  
for QGP production with the same value $p_0=2$ GeV as for the LHC.

Self-consistent screening of initial parton production can also be modelled 
in terms of a dynamical, medium-induced screening mass \cite{EMW}. 
The midrapidity partons are produced at times $\tau\sim 1/p_{\rm T}$, 
so the large-$p_{\rm T}$ partons are produced first, and partons with 
smaller $p_{\rm T}$ are produced later. In a medium of high-$p_{\rm T}$
partons, a dynamical  screening mass (electric, static) $m_g$
\cite{BMW92} is generated, and this mass then screens the production 
of partons with smaller $p_{\rm T}$. If the density of produced partons 
is high enough, {\it i.e.} if the $\sqrt s$ and $A$ are large enough, 
the screening  mass grows fast enough, causing a saturation of the 
parton cross section already in the perturbative regime 
$p_{\rm T}\gg\Lambda_{\rm QCD}$.  In practice, since the transverse part 
of the screening is not known,  we have  modelled the screening by simply 
making the replacement $\hat t(\hat u) \rightarrow \hat t(\hat u)-m_g^2$ 
in the partonic cross sections. Saturation in this approach 
coincides with Eq. (\ref{saturation}), and, by computing the first moment of 
the $p_{\rm T}$-distributions, one also concretely observes how the 
bulk of transverse-energy production is obtained from $p_{\rm T}\ge 2$~GeV 
at  the LHC (see \cite{EMW}).

To conclude this section, I will discuss nuclear parton distributions.
In the computation presented above, these are approximated as  
$f_{i/A}(x,Q^2) = Af_{i/p}(x,Q^2)$. Clearly, this gives a fair first 
estimate. It is, however, experimentally known that there 
are nuclear effects to the parton distributions. In the DIS measurements 
\cite{EMC,SLAC,NMC}, the ratio $F_2^A(x,Q^2)/F_2^D(x,Q^2)$ has been measured,
and four regions in $x$ can be distinguished: 
depletion at $x\lsim 0.1$ called ``nuclear shadowing'',
enhancement at $0.1\lsim x\lsim 0.3$ called ``antishadowing'',
depletion at $0.3\lsim x\lsim 0.7$ called ``emc-effect'', and
cumulative enhancement at $x\rightarrow 1$ (and beyond) called 
``Fermi motion''. See Ref. \cite{ARNEODO} for an extensive review 
of the experimental data and the various theoretical models.

\begin{figure}
\vspace*{-4cm}
\centerline{ \epsfxsize=6cm\epsfbox{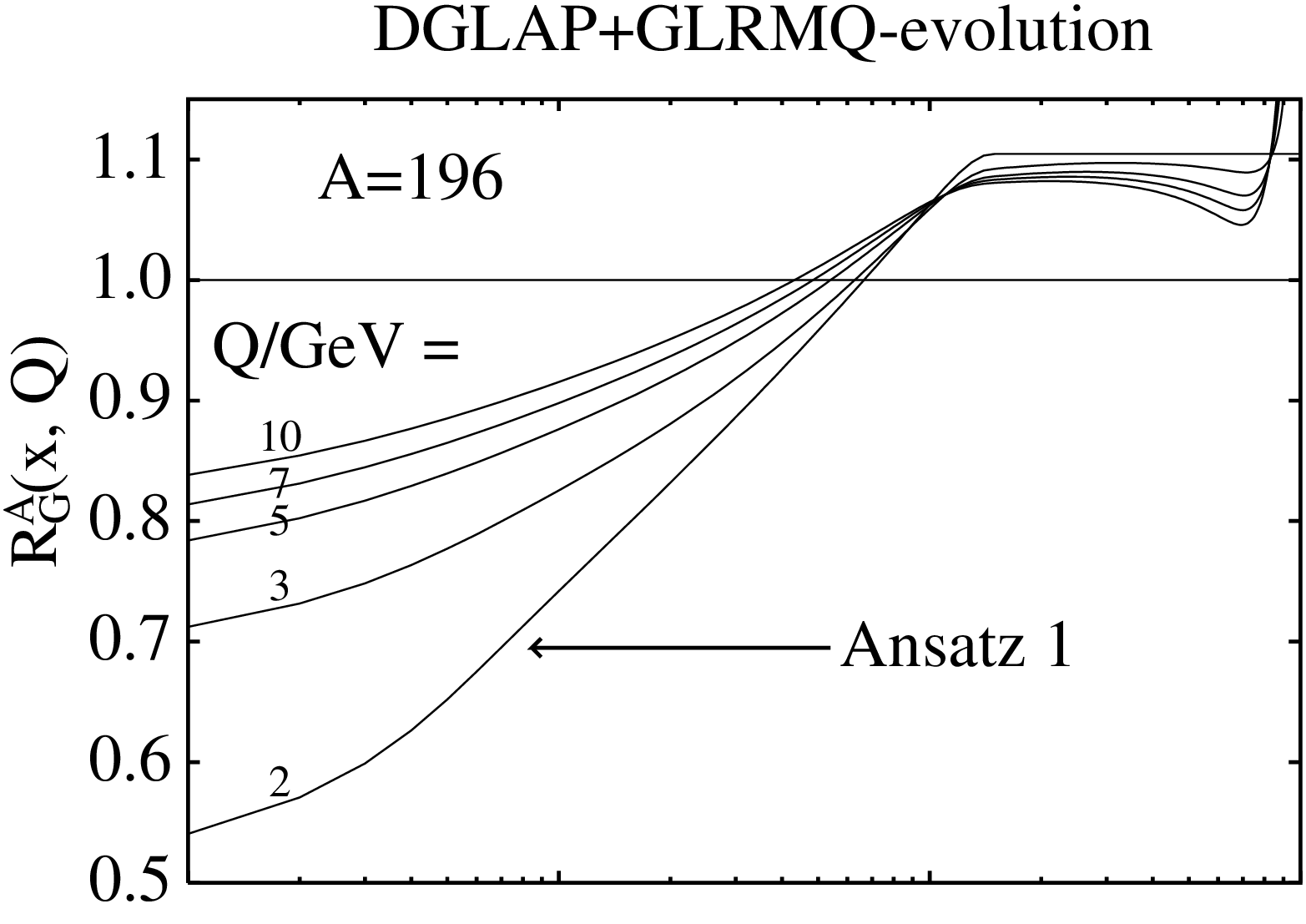}}
\vspace{-3.6cm}
\centerline{ \epsfxsize=6cm\epsfbox{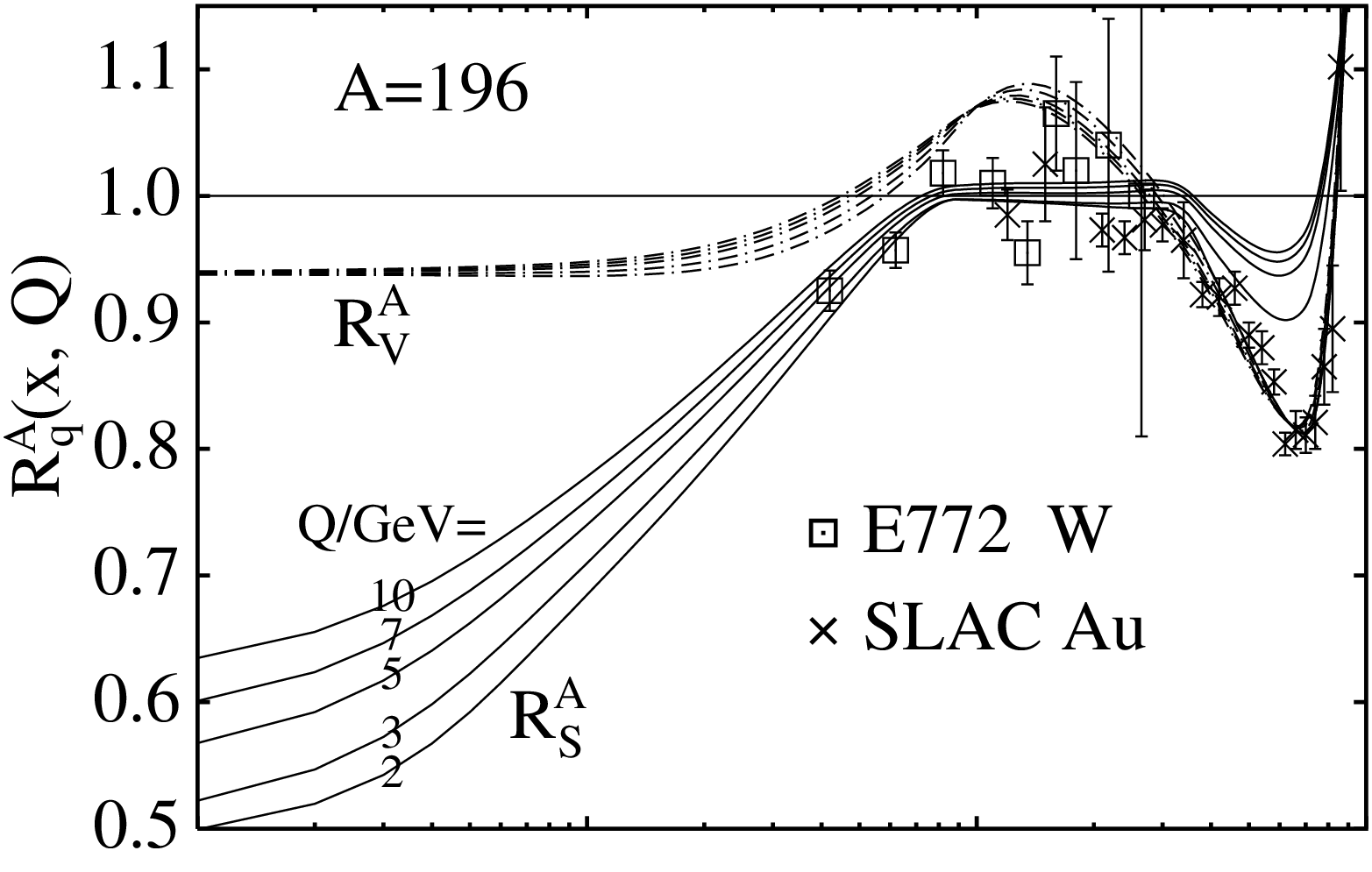}}
\vspace{-3.6cm}
\centerline{ \epsfxsize=6cm\epsfbox{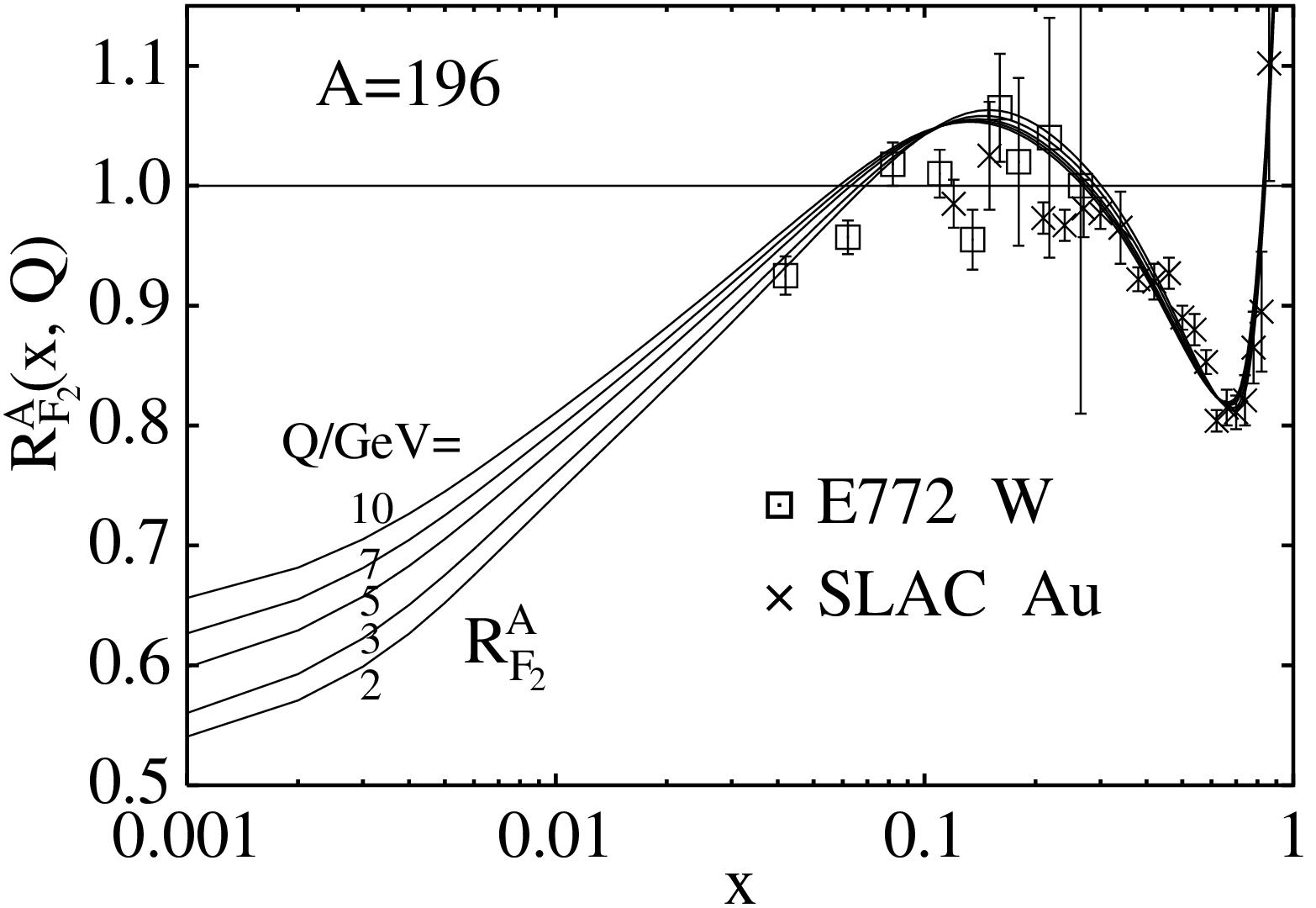}}
\vspace*{1cm}
\caption[a]{{\small
The ratios 
$R_i^A(x,Q^2) \equiv f_{i/A}(x,Q^2)/Af_{i/p}(x,Q^2)$ of
$i = g,\, V (= u_V+d_V),\, S (= \sum [q_S+\bar q_S]),\, F_2$ for 
$A=196$, as functions of $x$ at scales
$Q= 2, 3, 5, 7, 10$ GeV \cite{KJE93}. The evolution is lowest order DGLAP,
modified with the GLRMQ terms, and the data shown are from \cite{E772,SLAC} }

}
\la{shadow}
\end{figure}

For the mid-rapidity minijet production, practically only the shadowing 
region is relevant. It is experimentally observed that the $F_2$-ratio 
does not strongly depend on the virtuality $Q^2$, so a scale-independent 
ratio may give a first estimate of the nuclear effects to the nuclear 
quark and  antiquark distributions. However, as shown in \cite{KJE93}, 
it is not necessarily so for the gluons, but the QCD scale evolution 
\cite{DGLAP} should be taken into account in more detail. In 
Fig.~\ref{shadow}, I show the evolution of ratios 
$R_i^A(x,Q^2)= f_{i/A}/Af_{i/p}$ separately for gluon, valence- and 
sea-quark distributions, and for $F_2$, from \cite{KJE93}. In this 
analysis, charge and momentum sum rules are incorporated with the DIS 
data \cite{NMC,SLAC} and the dilepton data \cite{E772}. A reanalysis 
with  more modern parton distributions  is being prepared \cite{EKoR97}. 
It is also becoming possible to extract the gluon distributions from the 
logarithmic derivatives of  $F_2^{A_1}/F_2^{A_2}$ \cite{PIRNER}, so 
further constraints for the badly known nuclear gluon distributions are 
available. The scale evolution of nuclear parton densities has also been 
studied by other people, see Refs. \cite{KUMANO}. Finally, the role of 
the GLRMQ correction terms \cite{GLRpr,MQ87} (which were included in 
\cite{KJE93}) to the DGLAP equations \cite{DGLAP}  should also be 
considered in more detail, perhaps along the lines in  Ref. \cite{EQW}, 
but in the light of the most recent HERA-data \cite{HERAlatest}. To get 
an idea of the magnitude of the shadowing effects  in minijet production 
at the LHC and RHIC, I refer the reader to the computation  in \cite{KJE93}.

\section{Initial conditions for QGP at $\tau=0.1$ fm/$c$ from pQCD}

As described in the previous section, we obtain a first estimate of 
the initial conditions of QGP production by fixing the 
minimum $p_{\rm T}$ as $p_0= 2$ GeV. In this way, we describe the initial 
conditions at $\tau\sim 1/p_0=0.1$ fm$/c$. The predictions for the 
average numbers and transverse energies of partons produced into the
central rapidity unit in central Pb--Pb collisions at RHIC and LHC 
energies are summarized in Table 1 (with even too many decimals). 
Also individual contributions from  gluons, quarks and antiquarks 
are indicated \cite{EKR}.
There are five quite straightforward but important observations: 
\bigskip

\begin{table}
\center
\begin{tabular}{|c|c|c|c|c|c|}
\hline
      & $|y|<0.5$                    & total & $g$   & $q$    & $\bar q$   \\
\hline
LHC:  & $\bar N_{\rm PbPb}$          &4740   & 4349  & 199.9  & 191.2      \\
      & $\bar E_{\rm T}^{\rm PbPb}$  &14170  & 12960 & 620.8  & 588.8       \\

\hline 
RHIC: & $\bar N_{\rm PbPb}$          & 120.6  & 99.58 &12.96  & 8.102     \\
       & $\bar E_{\rm T}^{\rm PbPb}$ & 320.7  & 262.8 &36.06  & 21.87     \\

\hline
\end{tabular}
\caption[1]{{\small
The average numbers and transverse energies of semihard partons at 
$\tau=0.1$ fm$/c$ with $|y|\le 0.5$ and $p_{\rm T}\ge 2$ GeV in 
central Pb--Pb collisions, as given by Eqs. (\ref{NAA}) and (\ref{ET}). 
No shadowing nor $K$-factor is included and
the GRV-LO \cite{GRV94} parton distributions are used.}}
\la{table}
\end{table}

\begin{itemize}
\item 
Gluons strongly dominate the perturbative parton and 
transverse energy production: the initial parton system at $\tau=0.1$ fm/$c$
is about 90\% glue in the LHC and 80\% in the RHIC, so that at early times
the QGP is actually gluon plasma to the first approximation.

\item 
As discussed in Sec. 2, the parton system at the LHC is transversally 
saturated at $\tau=0.1$ fm/$c$, within the central rapidity unit. 
This means that production of more (softer) partons increases mainly the 
rate of final state fusions, not so much the transverse energy production. 
At the RHIC the same will happen but at a somewhat later stage which may 
not be entirely controllable by means of pQCD. In any case, it is  
demonstrated that the parameter $p_0$ acquires  a {\it dynamical} 
significance  in the URHIC.

\item 
By assigning baryon number 1/3 ($-$1/3) for each quark (antiquark) 
produced, and by approximating the volume of the 
parton system by $V=\pi R_{\rm Pb}^2\Delta y/p_0 = 13.4$ fm$^3$ for
Pb--Pb, the initial net baryon number density at $\tau=0.1$ fm/$c$
becomes $n_{B-\bar B} = 0.21\, {\rm fm}^{-3}$ for the LHC, and, 
$0.12\, {\rm fm}^{-3}$ for the RHIC. Thus, even at these ultrarelativistic
energies the initial net baryon number density is comparable to the 
nuclear matter density, $0.17\, {\rm fm}^{-3}$, even beyond it 
at the LHC.\footnote{Note that by the time $\tau=0.1$ fm/$c$, 
the Lorentz-contracted nuclear disks (the hard parts) are already receding; 
the transit time of the nuclei is  
$\tau_{\rm T}\sim 2R_A/\gamma\sim 5\times10^{-3} $ fm/$c$ for the LHC and 
0.1 fm/$c$ for the RHIC.} 
Of course, the high initial net baryon density  will dilute quite fast,
but it is interesting that the pQCD computation  gives such a high
initial density.  One would perhaps expect more baryon stopping at 
RHIC than at LHC. I should emphasize that the number presented here is 
only the perturbative part of net baryon number and that the increase 
for the LHC is entirely a gluon distribution effect because
the perturbatively produced net quark number in the central 
rapidity region is mainly due to valence quark--gluon scattering. 
The typical values of $x$ probed at the LHC are smaller than 
at the RHIC, so, because of the rapid rise of the  gluon densities, 
there are many more  gluons for the valence quarks to scatter with 
at the LHC than at the RHIC.

\item
To what extent can the initial system be considered thermalized?  
One immediately observes that there are far too few quarks and antiquarks
as compared to gluons, so that the system cannot be in chemical equilibrium.
How about the gluons alone then? In an ideal thermal gas of massless bosons 
the energy/particle is determined by $\epsilon_g/n_g=2.7 T_{\rm eq}$. 
From the numbers for the LHC  in Table \ref{table} we can determine that 
\begin{equation}
{\bar E^g_{\rm T}\over \bar N_{\rm PbPb}^g}
= \frac{\epsilon_g^{\rm pQCD}} {n_g^{\rm pQCD}}
= 2.98\,{\rm GeV}.
\end{equation}
In an ideal gas of massless gluons in complete thermal 
equilibrium,  the temperature can be computed from
$ \epsilon_g^{\rm ideal}=3\pi^2/90\cdot 16T_{\rm eq}^4 $, and for an 
ideal gas with an energy density 
$\epsilon_g^{\rm ideal}=\epsilon_g^{\rm pQCD}$
we find $T_{\rm eq}=1.10$ GeV. We see that
\begin{equation}
\frac{\epsilon_g^{\rm pQCD}}{n_g^{\rm pQCD}}  \approx
\frac{\epsilon_g^{\rm ideal}}{n_g^{\rm ideal}} = 2.7 T_{\rm eq}.
\end{equation}
So, as far as the energy per particle is concerned, the gluon system
is ``thermalized'' from the beginning at the LHC. At the RHIC, this is 
likely to happen somewhat later. One should, however, keep in mind 
that we did not consider isotropization at all here, and, that some 
uncertainty is connected with the assumption on an isotropization time.
In order to estimate this time, a more detailed space-time picture 
of initial parton production  is needed. Such a modelling is presented 
{\it e.g.} in Refs. \cite{EW94,PCM}.

\item 
Finally, let us consider the initial net baryon-to-entropy ratio 
of the early QGP. For a thermal boson gas $s=3.60n$, where $n$ is the 
number density of thermal gluons, the total initial entropy (glue only) 
is $S=15900$ for the LHC; so initially, at $\tau\sim 0.1$~fm/$c$, 
the net baryon-to-entropy ratio is $(B-\bar B)/S\sim 1/5000$. 
For the RHIC it will be larger, about 2/1000. In the further evolution 
of the QGP, this number will increase somewhat
due to the non-perturbative net baryon production; we estimated
the final $(B-\bar B)/S\sim8\times10^{-4} (9\times10^{-3})$ for the LHC 
(RHIC)  \cite{EK96}. Even though we are studying the same phase transition 
as took place in the  early Universe, we are still relatively far away 
from those extreme conditions regarding $(B-\bar B)/S$; there, the inverse 
of the specific entropy  is $\sim 10^{-9}$.

\end{itemize}

\section{Beyond the factorized minijet approach}

The simple approach of Sec. 2 without shadowing effects in the parton 
distributions  will fail at large rapidities when sufficiently high 
cms-energies and large nuclei are involved (see \cite{EK96}).  
Minijet production at large $y$ is a consequence of a  few large-$x$ 
partons scattering  against a large number of small-$x$ partons. 
Due to the rapid rise of the gluon distributions at small values 
of $x$, and without shadowing, the parton luminosity is high and  
the cross sections at large $y$ are not suppressed enough for the 
assumption of independent scatterings to hold. Eventually,  since 
$T_{AA}\sim A^{4/3}$, one will run into trouble in 
conserving energy  and baryon number (which scale as $\sim A$) 
for a large enough  $A$ and $\sqrt s$. In this case, coherence effects 
should be considered. However, within the central rapidity unit, 
where our focus is, the assumption of  independent parton scatterings 
still works quite well:
only energy $\sim {\cal O}(N_{AA}\times 2p_0) \ll A\sqrt s$ is consumed, 
and only a fraction of the available number of partons will scatter. 
Naturally, the latter criterion also serves as a further constraint for
the value of $p_0$ in URHIC. I should also emphasize that the approach 
will work even better once  nuclear (gluon) shadowing is included.

The question of the validity of factorization in minijet production in 
URHIC is, however, an important one, and additional production mechanisms 
should  be studied. During the recent years, quite a different approach 
to minijet production has been developed, not based on factorization.
McLerran and Venugopalan \cite{McLV} have suggested a model 
where a large colour charge density of a large nucleus, travelling along 
the light-cone, generates  a gluon field that is effectively classical. 
The idea then is that the gluon distribution function of the nucleus 
(perhaps even hadrons)  could be computed in the region 
$\Lambda_{\rm QCD}\ll k_{\rm T} \ll \mu$, where $k_{\rm T}$ is the 
transverse momentum of gluons, and the scale $\mu^2$ is the area-density 
of gluons  per unit rapidity.  Connection of this approach to evolution 
equations, especially to the BFKL \cite{BFKL}, has been considered in 
\cite{LEONIDOV}. The actual collision of two nuclei has been formulated 
in \cite{KOVNER}, and connection of the classical approach to Feynman 
diagrams has also been recently studied \cite{RISCHKE}. Also the 
connection to the BFKL-type minijet formula \cite{GLRpl,ELR96}
is under investigation \cite{GMcL}. It remains to be seen 
what the predictions of the gluon field approach eventually are in 
actual numbers for URHIC, and how well the gluon distributions and 
their QCD evolution  in nuclei and hadrons are accounted for.
In my opinion, it is very important to pursue this work into the direction
outlined in \cite{GMcL}, where the applicability of the model to URHIC 
at the LHC and RHIC was studied in more detail. 
Ultimately, one could hope that while collinearly factorized minijets 
dominate parton production at  $p_{\rm T} \gsim p_0 = 2$ GeV, the region of 
$\Lambda_{\rm QCD}\lsim p_{\rm T} \lsim p_0$ (where also non-linear effects 
will eventually  become important) could become better under control 
in terms of the  BFKL and/or gluon field approach.

Minijets can also be emitted from a BFKL ladder \cite{GLRpl}\footnote{
See the useful lecture notes by Del Duca \cite{DELDUCA} for a 
derivation of the basic concepts and for original references.}.
By assuming that  the small-$x$ increase  in $F_2$ is {\it entirely} 
due to the BFKL physics \cite{BFKL},  the {\it maximum} transverse 
energy deposit in the central rapidity region due to the minijets 
from a BFKL ladder can be studied. This was done in \cite{ELR96}. 
Since the increase in $F_2$ takes place only at $x\lsim 0.01$, the 
BFKL mechanism is not expected to be important in the RHIC nuclear 
collisions.

Minijet production without high-$p_{\rm T}$  tagging jets 
requires an introduction of unintegrated parton densities,
which evolve according to the BFKL equation \cite{BFKL}
(assumed to be homogeneous \cite{AKMS}), and which can be 
normalized to the ``known'' gluon distribution $xg(x,Q)$.
Effectively, minijet production from a BFKL ladder can be considered 
as an ${\cal O}(\alpha_s)$, $2\rightarrow1$ process where 
two virtual gluons fuse and form the minijet in the mid-rapidity. 
Technically, this minijet is just one fixed rung of the BFKL ladder 
(the emitted gluons of which are strongly ordered in rapidity). 
In \cite{ELR96}, we were unable to fix the overall normalization 
of the process, in the absence of any perturbative Born-level process 
to compare directly with.  By comparing the BFKL computation with the 
factorized jet-production  cross section (in the lowest order) 
at high $p_{\rm T}$, we were, however, able to argue that the 
transverse energy production from the BFKL minijets is still 
subleading at the LHC energies. 

Since the HERA  results for the increase of $F_2$ at small $x$
can be explained  by the leading $\log(Q^2)$ \cite{DGLAP} and/or the 
leading $\log(Q^2)\log(1/x)$ \cite{BF} approximations, 
the leading $\log(1/x)$ BFKL contribution is obviously 
{\it not} the dominant mechanism at the present values of $x$. Thus, my 
conclusion at this point is that the BFKL minijets certainly 
bridge the way towards softer physics at $p_{\rm T}<p_0\sim 2$ GeV, 
but the initial conditions relevant for the early QGP formation in 
the LHC nuclear collisions  seem to be  dominantly given by the 
minijets computed in collinear factorization. 

\section{Further evolution of the minijet plasma}

It is very important to understand when and how well the early
parton system can be described in terms of hydrodynamics \cite{BJ}. 
Not only the thermal signals, but also global variables 
such as final transverse energy and total multiplicity 
will strongly depend on the onset of pressure and flow effects.  
Normally, in hydrodynamical calculations the measured final-state 
hadron spectra are used for getting an estimate of the initial 
conditions (see {\it e.g.} \cite{VJGR}). These then, by definition, 
depend on the equation of state used. Semihard parton production
could now provide the hydro codes with additional and independent 
information on the early initial conditions. Of course, validity of 
applying hydrodynamics to the early  QGP  
ultimately depends on how completely and how fast the initial parton 
system  reaches a state of local thermodynamic equilibrium. 

I have explicitly shown by using pQCD, how thousands (hundreds) 
of partons, mainly gluons, will be produced within 
$\tau\sim0.1$ fm/$c$  into the central rapidity unit of URHIC at the  
LHC (RHIC). For the LHC, the perturbative gluon system is shown to be 
transversally  saturated  and thermalized in the energy/particle sense.  
This indicates that already the early evolution of the QGP in the LHC 
could perhaps be described in a simple hydrodynamical approach \cite{BJ}.
Initially produced perturbative partons do not necessarily, however, have
boost-invariant rapidity distributions \cite{EK96}. By making an 
extreme assumption of having a locally thermal (gluon) system
at $\tau\sim 0.1$ fm/$c$, and by using the rapidity distributions 
as computed in \cite{EK96} for the  initial conditions, it is
possible to study the deviations from a boost-invariant Bjorken 
picture. In \cite{EKR97} it is shown that the arising pressure gradients 
generate somewhat faster cooling of the (Q)GP as in the Bjorken picture.

At the RHIC also the soft component in particle and transverse energy 
production should be taken into account. 
A possible modelling for this as a source term in
Bjorken hydrodynamics, on top of the minijet initial conditions,
was suggested in \cite{EG93}, and also effects of dissipation 
and colour conductivity were studied. More complicated 3-dimensional 
systems with minijet initial conditions and 
density fluctuations have been considered in \cite{GR97}.

Related both to the initial conditions and to the further evolution 
of the QGP,  a more difficult and still open question is 
thermalization  of quarks, {\it i.e.} how fast, if at all, 
quarks and antiquarks come into chemical equilibrium with  
gluons \cite{BDMTW}.  Phenomenologically, this subject is 
vital for the thermal electromagnetic  signals \cite{VR}. 
Theoretically, the issue is space-time dependent phenomena 
in gauge theories, a virtually uncharted territory.

\section{Conclusions}

So far, the diverging minijet cross sections of Sec. 2 can 
be regulated in URHIC either by a dynamically determined cut-off parameter 
$p_0$,  or by modelling in a screening mass. These are plausible 
but phenomenological  considerations. Unless more theoretical 
progress in understanding  pQCD parton scattering in the 
few-GeV range is made, these will  remain so.  A quantitative study 
of the NLO terms in the  minijet cross sections, especially  
in transverse energy production,  will certainly give us a 
better handle on the validity of pQCD at these scales. 
A resummation, as in the Drell-Yan dilepton case \cite{AEGM},
will perhaps become possible eventually.
As discussed in Sec. 2, the nuclear gluon distributions 
are also poorly known; ultimately they should be measured better.
Coherent scattering and higher-twist effects should be studied in 
more detail, at large rapidities especially.

The main results given in  Secs.  2 and 3 are obtained by using 
collinear factorization and independent parton--parton scatterings.  
Minijets from the classical gluon field approach and from the
BFKL ladder discussed in Sec. 4 are not based on factorization. 
By studying the upper limit of minijet production from the BFKL ladder,
we were able to argue \cite{ELR96} that the BFKL-mechanism should 
still be subleading in $E_{\rm T}$-production at the central rapidities, 
at least up to the  LHC energies. Connection of the gluon field 
and BFKL approaches with the collinearly factorized minijet 
production should still be understood better \cite{GMcL}.

To conclude, it is clear that the analyses of semihard parton production 
discussed in this article can, and will be, sharpened  in various ways. 
I have, however, no doubt that the  semihard partons will play a major
role in  the formation and  evolution of the QGP in central rapidities 
during the first fractions of  fm/$c$ of the LHC and RHIC nuclear collisions.
Understanding the primary production mechanisms of partons will be the key to 
predicting and explaining the signals of the QGP and more global variables 
measured in the future heavy ion collisions at LHC and RHIC.

\bigskip
\noindent{\bf Acknowledgements.} 
I would like to thank J. Kapusta for the invitation to write this article, 
M. Gyulassy, A. Leonidov, L. McLerran and V. Ruuskanen for discussions, and
K. Kajantie for discussions and comments  on the manuscript.

\end{document}